\documentclass{article}
\usepackage{setspace}
\usepackage{abstract}
\usepackage[utf8]{inputenc}
\usepackage{color}
\usepackage{xcolor}
\usepackage{rotating}
\usepackage{graphicx}
\usepackage[normalem]{ulem}
\usepackage{amssymb}
\usepackage{gensymb}
\usepackage{xspace}

\font\fiverm=cmr5             \font\sevenrm=cmr7

          \font\sixrm=cmr6

\def\teq#1{$\, #1\,$}                         

\def\erg{\varepsilon}

\def\LEdd{L_{\hbox{\sevenrm Edd}}}

\def\lambdac{\lambda_{\hbox{\sevenrm c}}}

\def\taut{\tau_{\hbox{\fiverm T}}}  
\def\sigt{\sigma_{\hbox{\fiverm T}}}

\def\et3{\eta_3}
\def\th1{\theta_{-1}}
\def\r07{r_{0,7}}
\def\x05{x_{0.5}}

\def\cm{\hbox{~cm}}

\def\s{\hbox{~s}}

\def\MeV{\hbox{~MeV}}

\def\keV{\hbox{~keV}}

\def\erg{\hbox{~erg}}
\def\s{{\hbox{~s}}
\def\cm2{\hbox{~cm}^2}}

\def\T90{T$_{\rm 90}$}

\def\deltaD{\delta_{\hbox{\sixrm D}}}
\def\secheader#1{\vspace{5pt}\noindent{\bf #1}\hspace{3pt}}

\title{Rapid Spectral Variability of a Giant Flare from a Magnetar in NGC 253}

\author{O.~J.~Roberts$^{1,*}$, P.~Veres$^{2,*}$, M.~G.~Baring$^{3,*}$,\\ M.~S.~Briggs$^{2,4}$, C.~Kouveliotou$^{5,6}$,  E.~Bissaldi$^{7,8,*}$, \\ G.~Younes$^{5,6}$, S.~I.~Chastain$^{5,6}$, J.J.~DeLaunay$^{9}$, \\ D.~Huppenkothen$^{10}$, A.~Tohuvavohu$^{11}$, P.~N.~Bhat$^{2,4}$, \\ E.~G\"o\u{g}\"u\c{s}$^{12}$, A.~J.~van der Horst$^{5,6}$, J.~A.~Kennea$^{9}$,\\ D.~Kocevski$^{13}$, J.~D.~Linford$^{14}$, S.~Guiriec$^{5,6,15}$,\\ R.~Hamburg$^{2,4}$, C.A.~Wilson-Hodge$^{13}$, E.~Burns$^{16}$}

\begin{document}

\maketitle
\begin{itemize}

\item[1]{Universities Space and Research Association, 320 Sparkman Drive, Huntsville, AL 35805, USA.}
\item[2]{Center for Space Plasma and Aeronomic Research, University of Alabama in Huntsville, Huntsville, AL 35899, USA}
\item[3]{Department of Physics and Astronomy, Rice University, MS-108, P.O. Box 1892, Houston, TX 77251, USA}
\item[4]{Space Science Department, University of Alabama in Huntsville, Huntsville, AL 35899, USA}
\item[5]{Department of Physics, The George Washington University, 725 21st Street NW, Washington, DC 20052, USA}
\item[6]{Astronomy, Physics, and Statistics Institute of Sciences (APSIS), The George Washington University, Washington, DC 20052, USA}
\item[7]{Dipartimento di Fisica ``M. Merlin" dell’Università e del Politecnico di Bari, Bari, Italy}
\item[8]{Istituto Nazionale di Fisica Nucleare, Sezione di Bari, Bari, Italy}
\item[9]{Department of Astronomy and Astrophysics, The Pennsylvania State University, 525 Davey Lab, University Park, PA 16802, USA}
\item[10]{Center for Data-Intensive Research in Astronomy and Cosmology (DIRAC), Department of Astronomy, University of Washington Seattle, WA 98195, USA}
\item[11]{Department of Astronomy \& Astrophysics, University of Toronto, Toronto, Ontario, Canada M5S 3H4}
\item[12]{Sabanc\i~University, Faculty of Engineering and Natural Sciences, Istanbul 34956, Turkey}
\item[13]{Astrophysics Branch, ST12, NASA/Marshall Space Flight Center, Huntsville, AL 35812, USA}
\item[14]{National Radio Astronomy Observatory, P.O. Box O, Socorro, NM 87801, USA}
\item[15]{NASA Goddard Space Flight Center (GSFC), Greenbelt, MD 20771, USA}
\item[16]{Department of Physics and Astronomy, Louisiana State University, Baton Rouge, LA 70803 USA}
\end{itemize}

\begin{itemize}
\item[*]{Contact Authors: Oliver Roberts (oroberts@usra.edu), Matthew Baring (baring@rice.edu), Elisabetta Bissaldi (elisabetta.bissaldi@ba.infn.it) and P\'eter Veres (peter.veres@uah.edu).}
\end{itemize}

\begin{abstract}
\textbf{Magnetars are slowly-rotating neutron stars with extremely strong magnetic fields ($10^{13-15}$\,G~\cite{ThompsonDuncan1995,Kouveliotou1998}), episodically emitting $\sim100$\,ms long X-ray bursts with energies of $\sim10^{40-41}$\,erg. Rarely, they produce extremely bright, energetic giant flares that begin with a short ($\sim0.2$\,s), intense flash, followed by fainter, longer lasting emission modulated by the magnetar spin period (typically $2-12\,$s), thus confirming their origin~\cite{Hurley1999,Palmer2005}. Over the last 40 years, only three such flares have been observed in our local group \cite{Fenimore1996,Hurley1999,Feroci1999,Palmer2005}; they all suffered from instrumental saturation due to their extreme intensity. It has been proposed that extra-galactic giant flares likely constitute a subset~\cite{Duncan2001,Tanvir2005,Ofek2006,Mazets2008,Ofek2008} of short gamma-ray bursts, noting that the sensitivity of current instrumentation prevents us from detecting the pulsating tail, while the initial bright flash is readily observable out to distances $\sim 10-20\,$Mpc. Here, we report X- and gamma-ray observations of GRB 200415A, which exhibits a rapid onset, very fast time variability, flat spectra and significant sub-millisecond spectral evolution. These attributes match well with those expected for a giant flare from an extra-galactic magnetar~\cite{Yang2020}, noting that GRB 200415A is directionally associated with the galaxy NGC 253 ($\sim$3.5 Mpc away)~\cite{IPN2020}. The detection of $\sim3$~MeV photons provides definitive evidence for relativistic motion of the emitting plasma.  The observed rapid spectral evolution can naturally be generated by radiation emanating from such rapidly-moving gas in a rotating magnetar.}
  


\end{abstract}


On April 15$^{th}$ 2020 at 08:48:05.563746 UTC, the Gamma-ray Burst Monitor (GBM) onboard the \emph{Fermi} Gamma-Ray Space Telescope (\emph{Fermi}) was triggered by an extremely bright, short and spectrally hard event, initially classified as a short Gamma-ray Burst (GRB), GRB 200415A~\cite{Bissaldi2020}, which was also detected by several other instruments \cite{IPN2020,LAT2020,ASIM2020}. An offline search using time-tagged event (TTE) data from the Burst Alert Telescope (BAT) onboard the Neil Gehrels Swift Observatory (\emph{Swift}), obtained with the Gamma-ray Urgent Archiver for Novel Opportunities (GUANO)~\cite{Tohuvavohu2020} pipeline, also found the event. Using the light travel time of photons detected by the Inter-Planetary Network (IPN) of satellites, GRB 200415A was triangulated to a 17 square arc-minute region centered at RA and Dec. (J2000) of 11.88$^{\circ}$ (00h 47m 32s) and  -25.263$^{\circ}$ (-25d 15' 46"), respectively~\cite{IPN2020}. The relatively small error box of the localization significantly overlaps with, and, therefore, is highly suggestive of, GRB 200415A originating from the Sculptor Galaxy (NGC 253), an active star-bursting intermediate spiral galaxy located $\sim$3.5 Mpc away~\cite{Rekola2005}.

\begin{figure*}[ht!]   
\begin{center}
\includegraphics[width=\textwidth,trim=0 0 0 0,clip]{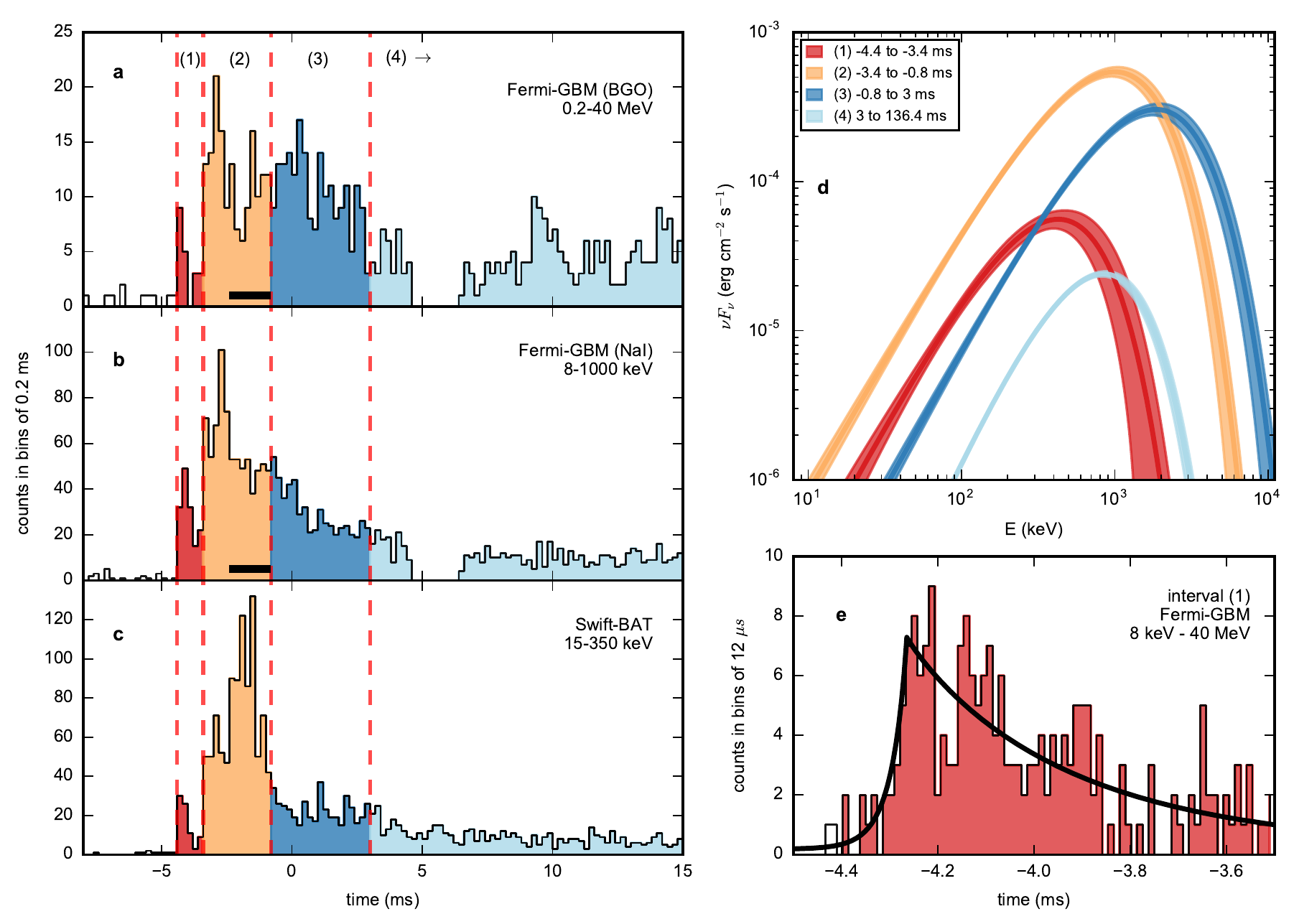}
\caption{\textbf{Temporal and Spectral Variability of GRB 200415A.} (Left:) Lightcurves with 0.2 ms resolution for a \emph{Fermi} GBM BGO detector (a), NaI detector (b) and \emph{Swift}-BAT (c). The BAT lightcurve was shifted by 5.7 $\mu$s to account for the light-travel time between the spacecrafts. Panel (d) shows the spectra for the four intervals. The shaded area indicates 1$\sigma$ confidence regions.
Using the BAT TTE data, we identify that the GBM TTE bandwidth (see Methods) was exceeded from -2.4~ms to -0.8~ms (horizontal black lines in panels a,b), resulting in a 47.3\% loss of flux in interval (2). 
There is also a $\sim$1.8~ms data gap from 4.6 to 6.4 ms caused by a CSPEC packet blocking the GBM TTE data, resulting in a 3.47\% loss in interval (4). Panel (e) shows the first pulse with high temporal resolution (12 $\mu s$) and the fitted pulse profile.
}
\label{fig:Fig1} 
\end{center}   
\end{figure*}

We use the BAT TTE data to determine the duration due to bandwidth saturation of the high-time resolution GBM TTE data (see Methods). We find the T$_{90}$ duration of GRB 200415A (the time interval over which 5\% to 95\% of the total counts was accumulated~\cite{Kouveliotou1993}), to be 140.8$^{+0.5}_{-0.6}$~ms (1$\sigma$). Correspondingly, the event T$_{50}$ duration is 54.7$^{+0.5}_{-0.4}$~ms (1$\sigma$). Further, our detailed temporal analysis of the event lightcurve shows that the rise time (10 to 90\%) of the first pulse is $T_{\rm rise}=77\pm 23 \mu s$ (1$\sigma$) (Fig.~\ref{fig:Fig1}, panel (e)).

We performed a timing analysis on the GBM light curve, searching for a rotational frequency in the range 0.02-50 Hz, finding no clear pulsation. We also searched the 40-4000 Hz window for Quasi-Periodic Oscillations (QPOs), possible signatures of seismic vibrations seen in the oscillating tails of confirmed Giant Flares (GFs); a candidate broad QPO was found at a frequency of $\nu\sim 180$~Hz in the decaying tail of GRB 200415A with $\sim 2.5\sigma$ significance (see Methods).

\begin{figure*}[ht!]   
\begin{center}
\includegraphics[width=\textwidth,trim=0 0 0 0,clip]{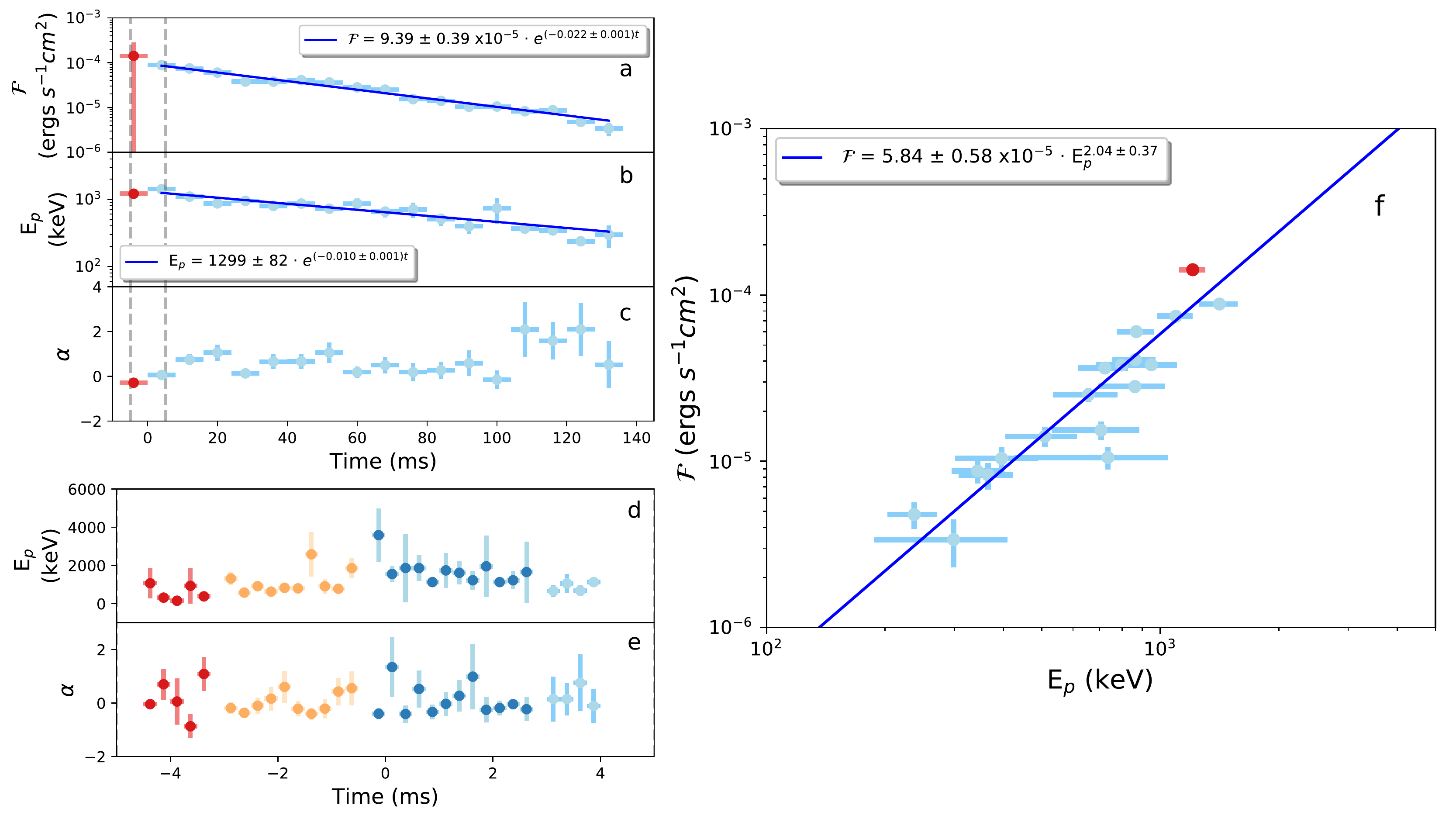}
\caption{\textbf{Flux and Spectral Evolution of GRB 200415A.} Spectral variability of GRB 200415A. Panels a, b and c show the energy flux or ${\cal F}$ (a), peak energy or E$_{p}$ (b) and $\alpha$ (c) evolution over the $T_{90}$ duration ($\sim$141~ms). $t$ is the time in seconds. The blue lines are exponential fits to ${\cal F}$ and E$_{p}$ over the tail emission of GRB 200415A (5~ms after zero time). This results in a decay timescale of 45 $\pm$ 3~ms for ${\cal F}$. The decay time for E$_{p}$ is 100 $\pm$ 1~ms. The temporal binning is 8~ms for these panels. Panels d and e show how the E$_{p}$ and $\alpha$ trend on sub-ms timescales ($\Delta$t = 250 $\mu$s), over the interval shown between the dashed grey lines in Panels a through c. Panel f, shows the relationship between ${\cal F}$ and  E$_{p}$ over parts of the burst not affected by data saturation, using 8~ms temporal resolution. The blue line is a power-law fit to this data (exclusively after the main peak), which shows ${\cal F}$ $\propto$ E$_{p}^{2}$. The spectral range used for these measurements was 8~keV to 10~MeV. The coloring scheme follows that in Fig.~\ref{fig:Fig1}. The zero time reflects the time of the GBM trigger. All fit errors are at the 1$\sigma$ confidence level. For more information on their derivation, see Methods.
}
\label{fig:Fig2} 
\end{center}   
\end{figure*}

We performed time-integrated and time-resolved spectral analyses of the GBM data, focusing on the sub-ms structures in the lightcurve, shown in Figs.~\ref{fig:Fig1} and \ref{fig:Fig2}. The very high rate might cause the electronic signals of photons to overlap (pulse pile-up), causing their energies to be incorrectly measured, and spectral distortions. We evaluated this effect in the brightest interval (2 in Fig.~\ref{fig:Fig1}) and determined that it was negligible, (see Methods). Among several spectral models used, we found that a power-law with an exponential cutoff (Comptonized) model, fit the data best; the Comptonized spectral parameters are presented in Table 1 (see Methods). The highest energy photons reliably associated with GRB 200415A, have energies of $\sim 3 \MeV$ (see Methods). Using time-resolved GBM spectral analysis with corrections from the BAT, we find a time-integrated isotropic equivalent energy output of $E_{\rm iso}=(1.51\pm0.021) \times 10^{46}$ erg (see Table~1). The peak isotropic luminosity is $L_{\rm iso,\max}= (1.53 \pm0.13) \times 10^{48} $~ erg s$^{-1}$, while the total event luminosity is $L_{\rm iso}= (1.07 \pm0.17) \times 10^{47} $~ erg s$^{-1}$. Our time-resolved spectral analysis shows remarkable sub-ms variations (Fig.~\ref{fig:Fig2}d,e) over a 10\,ms interval, encompassing intervals 1, 2, 3, and part of 4. In Fig. 2d, we notice that the peak energy ($E_{\rm p}$) reaches its highest value at the onset of interval 3, while it remains relatively constant throughout most of the event.

The photon index ($\alpha$), stays relatively constant at $\alpha \sim 0$ during the event, which would be highly unusual for a short GRB. Figs.~\ref{fig:Fig2}a,b show exponential decay trends in both energy flux ${\cal F}$ and $E_{\rm p}$ over interval 4, which is clearly discerned from the tail of GRB~200415A. The energy flux decay in Fig.~\ref{fig:Fig2} occurs on a timescale of $\tau$ = 45 $\pm$ 3~ms. The $E_{\rm p}$ is observed to decay on a longer timescale of $\tau$ = 100 $\pm$ 1~ms in the same figure. This exponential behaviour has been observed in other extra-galactic GF candidates~\cite{IPN2020}.
A distinctive ${\cal F} \propto E_{\rm p}^2$ correlation was discovered (Fig.~\ref{fig:Fig2}f); a signature of a relativistic wind. This unprecedented result is clearly observed in the GBM data for GRB~200415A, devoid of detector saturation effects. Such saturation effects likely precluded the ability to discern this trend cleanly from previous observations of galactic GFs from SGRs 1900+14 and 1806-20. 

Finally, we searched for radio emission associated with GRB 200415A, in four observations of the NGC~253 taken with the Karl G. Jansky Very Large Array (VLA)~\cite{JVLA}, 4.3 to 51.2 days after the event trigger. No significant variable or transient emission was identified. 

Previous studies postulated that about 1-20\% of sGRBs could be extra-galactic GFs \cite{Duncan2001,Palmer2005,Ofek2007,Hurley2011}.
The sample of galactic GFs is very small and their properties are ill-determined due to instrumental effects from their extreme intensity. We therefore, first compare the GBM observations of GRB~200415A to the GBM observations of short GRBs (sGRBs)~\cite{AvK2020}. We find that the 64~ms peak photon flux ($P_{64}^{\rm catalog}= 73.7\pm 2.1\,{\rm photons} \cm^{-2} \s^{-1}$) of GRB 200415A lies at the 97.5$^{th}$ percentile of the sGRB distribution, the peak energy ($E_{\rm p}^{\rm catalog}=998\pm45 \keV $) at the  79$^{th}$ percentile, and the photon index ($\alpha^{\rm catalog}= 0.39 \pm 0.09$) at the 88.5$^{th}$ percentile. It is similarly near the edge of the \teq{\alpha} distribution for the burst population detected with the Burst and Transient Source Experiment (BATSE) on the Compton Gamma-ray Observatory \cite{Kaneko-2006-ApJS}. Consequently, we find the flat, hard spectral slope, high $E_{\rm p}$ and peak flux during the brightest 64 ms of GRB 200415A (shown in Figs.~\ref{fig:Fig1} and \ref{fig:Fig2}) to be unusual for sGRBs, thus better explained as the initial spike of a magnetar GF from NGC~253. This interpretation is further motivated by similarities of the properties of this event to previously proposed extra-galactic GF candidates \cite{Mazets2008,Frederiks2007}. A rapid rise time is characteristic of a GF onset. Compared to the rise times reported in the GRB catalogs of GBM and BATSE, this rise time is considerably shorter than for any event in these samples, and is shorter than extreme examples of short variations, such as $\sim 100~\mu s$ for GRB 910711 \cite{Bhat+92tvar} and $2.8~ms$ for GRB 090228 \cite{MacLachlan+13tvar}.

Unfortunately, we could not detect the magnetar period-modulated tail with putatively $E\sim10^{44} \erg$. This `smoking gun' evidence for a GF, is observed over several hundreds of seconds in all three confirmed magnetar GFs, but is absent in GRB 200415A as it is likely below the detection threshold for GBM given its distance to NGC~253. For other extra-galactic GF candidates this feature is similarly undetected~\cite{IPN2020}.

The standard picture for the origin of the GFs is the release of energy triggered by fracturing the crust of the neutron star\cite{ThompsonDuncan1995} by large sub-surface magnetic fields, depositing hot plasma into the inner magnetosphere. Using a GF interpretation, the GBM observations indicate that the MeV-band emission must come from a relativistic outflow that is initially, highly `optically thick' (opaque). The enormous isotropic equivalent luminosity of \teq{L_{\rm iso}\gtrsim 10^{47}}erg $s^{-1}$ is orders of magnitude larger than the fiducial Eddington luminosity limit, \teq{\LEdd \sim 10^{38}}erg $s^{-1}$, for a neutron star of solar mass, $M_{\odot}$~\cite{Rybicki-1979}. This limit defines when radiation pressure associated with electron (Compton) scattering overwhelms gravity and pushes hydrogenic gas away from the surface to high altitudes. For GRB 200415A, we thus expect a relativistic wind~\cite{ThompsonDuncan1995} with bulk Lorentz factor \teq{\Gamma \gg 1} (i.e., speed $c (1-1/\Gamma^2)^{1/2}$) to be present, putatively over the magnetic poles.  At high altitudes, the radiation pressure abates and the wind ``coasts'' with constant $\Gamma$.  Transparency of the wind to quantum electrodynamical (QED) pair production $\gamma\gamma\to e^+e^-$ of electrons ($e^-$) and positrons ($e^+$) by 3 MeV photons, detected by GBM, unambiguously implies that $\Gamma > 6$ (see Methods). This is a more stringent bound than was possible for the $\lesssim 1$ MeV photons seen in the initial spikes of galactic GFs~\cite{Feroci1999,Boggs-2007-ApJ}. This GBM limit is consistent with the stronger constraints due to the detection of GeV photons by {\it Fermi}-LAT~\cite{LAT2020}. High Lorentz factors ($\Gamma \gtrsim 100$) are predicted from dynamical models~\cite{Meszaros+00phot} of thermal fireballs with high peak energies, usually applied in GRB emission modeling. The high opacity to QED {\sl magnetic} pair creation \teq{\gamma \to e^+e^-} in the inner magnetosphere of magnetars \cite{Baring-2001-ApJ} indicates that this wind is likely dominated by \teq{e^{\pm}} pairs, with limited baryonic content. The dense wind relativistically boosts its embedded radiation to higher frequencies via the Doppler effect and ``beams'' or collimates it into a radiation emission ``cone'' of opening angle \teq{\Theta_{\rm coll}\sim 1/\Gamma} radians. The observed correlation between the energy flux and the peak energy, ${\cal F} \propto E_{\rm p}^2$ (Fig. \ref{fig:Fig2}) can be readily explained by relativistic Doppler boosting (see Methods).

The GBM spectrum is very flat, i.e. has a relatively high value for its power-law spectral index $\alpha$. This is expected for a wind that for the most part is highly opaque to electron scattering, a so-called (ballistic) Compton cloud.  The huge \teq{L_{\rm iso}} indicates electron number densities of \teq{n_e \gtrsim 10^{29}}cm$^{-3}$ when the plasma is close to the stellar surface, for which the electron-photon Compton scattering mean free path is \teq{\lambda_s \sim 10^{-5}}cm or shorter. Opacity to scattering persists out to altitudes of \teq{R\sim 10^9-10^{10}}cm and higher. The very flat indices, \teq{\alpha \approx -0.2 \to 0.3} are similar to those identified in Lin et al.\cite{Lin-2012-ApJ} for normal SGR J1550-5418 bursts, wherein it was concluded that the high \teq{n_e} densities must seed rampant Comptonization that yields spectra beginning to approach a modified blackbody (Wien) form \cite{Rybicki-1979}. The same situation is anticipated for GRB 200415A, with its markedly higher deduced \teq{n_e} values. Yet, the radiation is not truly thermal: the photon energy flux at the source as inferred from the GBM-detected flux is substantially inferior to that appropriate for a Planck distribution from a relativistic wind that yields the same \teq{E_{\rm p}} in the observer frame (see Methods). This strongly contrasts the picture of GRB fireballs, which possess fluxes and $L_{\rm iso}$ many orders of magnitude larger, yet with similar \teq{E_{\rm p}} values. The broad, flat spectrum of the GF may in fact be a superposition of Comptonized Wien-like spectra from different altitudes spanning a range of effective temperatures in the adiabatically cooling wind. Note that such flat \teq{\alpha} indices are inconsistent with synchrotron emission scenarios commonly invoked for GRB spectral interpretation \cite{Preece-1998-ApJ}. 

The spectral evolutionary sequence displayed in Fig.~\ref{fig:Fig1} suggests that the initial hardening might reflect relativistically-boosted emission coincident with later stages of the wind acceleration epoch, followed by some cooling during the coasting phase of the laterally-expanding wind. Alternatively, it could be a convolution of the influences of stellar rotation and radiation beaming (collimation) into a ``cone'' subtending an angle of \teq{\Theta_{\rm coll}\sim 1/\Gamma}.  As the Doppler-boosted cone sweeps across our line of sight, the intensity would thus increase, peak and then decline, accompanied by a spectral hardening and subsequent softening; this is commensurate with the evolution displayed in Fig.~\ref{fig:Fig1}. Such a transient ``relativistic lighthouse beaming'' effect would generate the decay time \teq{\tau \approx 45\,}ms of the tail in Fig.~\ref{fig:Fig2} of a \teq{P=8}s rotation period for a $\Gamma\sim 30$ bulk flow (see Methods), noting the intrinsic coupling \teq{\tau \sim P/(2\pi \Gamma )} for the rotating beam.


\newpage

\newpage
\section*{Methods}
\label{sec:methods}

\secheader{GBM Observations and Data Processing}
\emph{Fermi}-GBM consists of 12 thallium-doped Sodium Iodide (NaI) detectors and two Bismuth Germanate (BGO) detectors. The uncollimated NaI scintillator detectors are clustered into four groups of three detectors at each corner of the spacecraft, arranged such that any cosmic source above the Earth's horizon will illuminate one cluster. The combined effective spectral range of both the NaI and BGO detectors are $\sim$ 8 keV to 40 MeV. GBM has several data types, which are either produced by a trigger or continuously. Triggered data types include CTIME data (binned to 64 ms resolution with 8 energy channels), and CSPEC data (binned to 1024 ms with 128 energy channels). The GBM TTE triggered data are the tagging of individual counts in the detectors within 5 minutes of the trigger time and have a relative timing resolution of $\sim$ 2 $\mu$s and 128 energy channels. The energy channels for all of these data types are pseudo-logarithmically spaced. More information on these and other data types, as well as other details pertaining to the instrument, can be found in Meegan et al.\cite{Meegan2009}. 

A bandwidth limit between GBM and the \emph{Fermi} spacecraft restricts the time-tagged GBM event rate to $\approx$ 375,000 events per second, summed over all detectors and energies \cite{von_Kienlin_2012}. When the GBM TTE rate exceeds this value, TTE counts are lost in a manner that is not biased by detector or energy (see Fig. \ref{fig:Fig1}), consequently spectral analysis to obtain spectral shape parameters is still possible, but the flux normalization will underestimated. During the saturated interval, sub-intervals may be weighted differently, but the data is still in correct order. Additionally the transmission of another GBM datatype to the spacecraft can block the TTE data, causing a gap. Other GBM data types are not effected by this bandwidth limit but have inadequate temporal resolution for this event. 

Due to high count rates, the electronic signals of the counts can overlap (pulse pile-up), causing incorrect energy measurements and spectral distortions. To check for this in the GBM data for GRB 200415A, we used an analytical method~\cite{Vandiver} that was verified by high-rate measurements with a GBM detector~\cite{highrates}. This method was applied to the spectral model and rate of interval~2, for the detector with the highest
flux (NaI 1). The spectral shape of the detector counts are only slightly modified by pulse pile-up, with a change in slope below 30 keV and a loss of higher energy counts starting at 200~keV, reaching a 5\% loss above 400 keV. We note that in joint fits, the lower-rate 
BGO data help constrain the results. Pulse pile-up is otherwise not included in our analyses.


The localization accuracy of an all-sky monitor like GBM is limited, especially for short events. While an initial position for GRB 200415A was promptly determined using GBM ~\cite{GCN27579,GCN27580} using several methods~\cite{Goldstein2020,Berlato2019}, each position had several degrees of error. We thus used the IPN localization\cite{IPN2020} to generate Detector Response Matrices (DRMs).

Each GBM detector was checked to see whether their viewing angles were within 60$^{\circ}$ of their respective on-axis position to the source, and whether they were blocked by parts of the spacecraft. From this analysis, we identified NaI detectors 0, 1, 2, 3 and 5 and the BGO 0 detector as having satisfied these criteria. Spectral parameters were determined by selecting the model that best fit the data, by looking at variations in the Cash-Statistic (C-Stat.) per degree of freedom (dof) \cite{Gruber-2014-ApJS}. A simple power-law, a power-law with an exponential cut-off (Comptonized), a black-body, a Band function \cite{Band+93} and combinations of these models were used. The values in Table 1 and in Figs. \ref{fig:Fig1} and \ref{fig:Fig2} are all spectral parameters derived form the Comptonized model fit to the data, which was found to be the best. 
The estimated covariance of the optimal values to the fits of the spectral parameters in  Fig.~\ref{fig:Fig2}a,b,f, use the minimization of the sum of the squared residuals. The diagonals provide the variance of the estimated parameter. One standard deviation error on the parameters was done by taking the square-root of the diagonals from the covariance matrix. The full fits to the data in Fig.~\ref{fig:Fig2}a,b,f, are ${\cal F}$ = (9.39$\pm$0.39) $\times$ 10$^{-5}$ $\cdot$ e$^{(-0.022\pm0.001)t}$, $E_{\rm p} = (1299\pm  82) \cdot e^{(-0.010\pm0.001)t}$ and ${\cal F}$ = (5.84$\pm$0.58) $\times$ 10$^{-5}$ $\cdot$ $(E_{\rm p}/1000~keV)^{2.04\pm0.37}$, respectively. ${\cal F}$ is measured in erg cm$^{-2}$ s$^{-1}$, $t$ in seconds and $E_p$ in keV. 

All of which include 1$\sigma$ errors, as shown in the Fig. \ref{fig:Fig2}. This results in decay times (with 1$\sigma$ errors) of 45 $\pm$ 3~ms and 100 $\pm$ 1 ms for  ${\cal F}$ and $E_{\rm p}$, respectively. All subsequent reported times are measured relative to the trigger time at 08:48:05.563746 UTC on April 15$^{th}$ 2020.

As a precursor was observed 142~s before the initial spike of the galactic GF from SGR 1806-20~\cite{Boggs-2007-ApJ}, we searched for precursor emission in a $2400 \s$ interval preceding GRB 200415A using the GBM {\tt targeted search} \cite{Goldstein+19targeted}: this search returned no candidates.

\secheader{\emph{Swift}-BAT Observations and Data Processing}
The BAT is a coded-mask imaging instrument with a 2 steradian field of view (FoV) in the 15-350~keV energy band \cite{Barthelmy2005}, but is capable of recording photons with energies up to $\sim500$ keV. For GRB 200415A, we used a new capability to provide data from BAT: the Gamma-ray Urgent Archiver for Novel Opportunities (GUANO) \cite{Tohuvavohu2020}, which allows the time-tagged BAT event data to be saved on demand, based on triggers from external instruments. At the time of the burst, BAT was oriented such that the source illuminated the detector at an angle of $48.51\degree$ from the center of the FoV. Although the location was entirely outside of the coded FoV (by a few degrees), the source was sufficiently bright to penetrate the graded-Z shield on the sides of the instrument and deposit significant flux onto the detector array. If the burst had been within the FoV, it would have resulted in a prompt autonomous trigger and most likely, would have saturated the instrument.

Using an instrument and spacecraft mass model (Swift Mass Model: SwiMM  \cite{Sato:2006zz}) in GEANT4 and Monte Carlo methods, we generated custom DRMs for the  orientation of the source, thus allowing the use of XSPEC  \cite{Arnaud96XspecProc} to fit spectra and derive fluxes from the count rates. The effective area response of the BAT in this orientation is both substantially reduced and altered with respect to sources originating from within the FoV, due to the shielding. In this orientation, the BAT effective area peaks at ~200 keV with an effective area of $\sim400$ cm$^{2}$, quickly dropping to $<50$ cm$^{2}$ at around 50 keV. 

The BAT data were also used for timing analysis, such as the $T_{90}$ and $T_{50}$ durations, shown in ED Fig.~1. The errors for this measurement were estimated by taking a standard deviation of the median fluence value after the cumulative counts curve flattened. 


\secheader{Rise time}
We perform an analysis of the first pulse using the GBM TTE data from the NaI and BGO detectors (interval (1), Fig. \ref{fig:Fig1}(e)), that starts at $-4.4$ ms and is unaffected by data saturation. 
We fit a pulse function introduced by Norris et al. \cite{Norris+96pulse} to model the properties of GRB pulses. The count rate as a function of time is described by $I(t)\propto \exp(-((T_{\rm peak}-t)/T_{1})^\nu)$ before the peak of the lightcurve and $I(t)\propto \exp(-((t-T_{\rm peak})/T_{2})^\nu)$ after the peak. We find $T_1=32.8\pm 9.9 \mu s$. The rise time, defined as the time elapsed between the 10\% and 90\% of the peak yields $T_{\rm rise}=77\pm23 \mu s$ at the 1$\sigma$ confidence level.

\secheader{Energetics and Time-resolved Spectra}
We divided the lightcurve into 4 intervals (see Fig.~\ref{fig:Fig1} and Table \ref{tab:Table1}). The differential photon number spectrum ($dN/dE$) in all four intervals is best described by a power law with an exponential cutoff. The cutoff is parametrized as the peak energy of the $\nu F_\nu$ spectrum, $E_{\rm p}$. 
The first interval (1) contains the first pulse, present in both GBM and BAT. It is a `clean' pulse, not affected by saturation and there is no overlap with other pulses. Interval (2) is the brightest part of the lightcurve, affected by TTE saturation in GBM. Interval (3) is the hardest with peak energy $E_{\rm p}=1.9\MeV$. The fourth interval has a featureless decay out to 136.4 ms and contains most of the fluence. The spectral fits for intervals (1), (3) and (4) are illustrated in ED Fig.~2. The comparison of the energetics to sGRBs were over 64~ms, a timescale generally reported with sGRB properties in mission catalogues.

\secheader{Highest Energy Photons}
ED Fig.~3 shows the individual TTE counts in GBM BGO detector 0. The GBM detectors register energy deposits and are unable to identify particle type, nor can they determine photon arrival direction. Therefore, it is impossible to determine with certainty whether any particular TTE is from GRB 200415A, another gamma-ray source or is due to other background. Instead, we can determine whether a rate increase is statistically significant and therefore associated with GRB 200415A. We use a Bayesian method applicable to Poisson rate data, for the classic on-source / off-source method of source detection~\cite{Gregory}. The method uses data from two time intervals, an off-source or background interval and an on-source interval, to test two hypotheses: 1) that all of the TTE are due to background, and 2) that there are excess TTE above background in the on-source interval due to the source. The method needs a prior expectation for the source count rate, which we obtain from the spectral fit in the on-source interval. For the 2.5 to 3.5~MeV energy range, there are 100 TTE in a 0.99~s background interval and 9 TTE in a 6.4~ms on-source window (blue box in ED Fig.~3 The calculated probability for a source signal (GRB 200415A) above the background is 0.9999997.  This energy range is  well above the threshold for $\gamma\gamma$ pair creation; the two TTE with the highest energies are 3.0 and 3.1~MeV.

We also consider a higher energy range, 3.5 to 10~MeV. In this energy range  there are 88 off-source TTE and three on-source TTE (red box in ED Fig.~3) with energies of 4.0, 6.7 and 8.8 MeV. The probability that the three on-source TTE are an excess rate that should be attributed to GRB 200415A is 0.966. Consequently, we do not consider these three TTE sufficiently significant to use for our Lorentz factor analysis.

The definitive detection of $E_{\rm max}\sim3\,$MeV photons for GRB 200415A by GBM is at an energy well above the pair-production threshold. Dedicated gamma-ray instrument observations of the main pulse of galactic GFs have typically been saturated, and likely suffer from spectral distortions (pulse pileup). Thin particle detectors have reported usable spectra (e.g., the WIND SST silicon detectors measured a spectrum for the December 27, 2004 GF main pulse of SGR 1806-20 extending to 1~MeV \cite{Boggs-2007-ApJ}). Gamma-ray detector observations of extra-galactic GFs will likely provide the best results on their spectra, obtaining excellent statistics without spectral distortions due to extreme fluxes.

\secheader{The Environment of the MeV Emission Region}
The appearance of emission above the two-photon pair creation
(\teq{\gamma\gamma\to e^+e^-}) threshold of $m_ec^2 =511$ keV in
GRB 200415A can be used to provide a lower bound to the wind bulk Lorentz
factor \teq{\Gamma} relative to the magnetar.  This is a traditional
practice in GRB studies \cite{Krolik-1991-ApJ,Baring-1997-ApJ,Granot-2008-ApJ}, 
using the bulk motion to reduce the pair opacity through relativistic collimation of the radiation field within an angle \teq{\Theta_{\rm coll}\ll 1}. The spectrum of this giant flare with its emission limited to energies
$E<E_{\rm max}\sim 3\,$MeV demands an individualized calculation of pair
opacity. The most conservative estimate corresponds to \underline{all}
observed GBM photons being below the 511 keV threshold in the comoving
frame of the plasma/photon gas.  Then one simply has $\Gamma > E_{\rm
max}/m_ec^2 \sim 6$. Somewhat higher values can be obtained with
additional assumptions of source spectral extension beyond $E_{\rm
max}$, albeit undetectable due to the radiation background. The pair opacity
constraint on $\Gamma$ is more restrictive than is possible for $E_{\rm
max}\sim 1\,$MeV values appropriate for the SGR 1900+14 and SGR 1806-20 
giant flares.  Expectations from isotropic fireball dynamics
modeling \cite{Nakar-2005-ApJ} and especially Fermi-LAT observations \cite{LAT2020} suggest much larger values for the bulk $\Gamma$.

The GBM emission is non-thermal. For a distance $d=3.5\,$Mpc to the
Sculptor galaxy, the detected photon energy flux can be mapped over to an
energy flux ${\cal F}(R)$ through a surface located at a distance $R$
from the magnetar's center. The entries in Table~\ref{tab:Table1} for
interval (3) indicate that ${\cal F}(R)\sim 2 \times 10^{30}
(10^8\hbox{cm}/R)^2$ erg cm$^{-2}$s$^{-1}$. If this radiation were to
be thermal, then this flux should be comparable to the Doppler-boosted
Planck spectrum that generates the energy peak value $E_{\rm p}\sim
1856\,$keV for this interval. Suppose the Planck spectrum in the wind
frame has some temperature $T_{\rm w}$. For a wind moving with speed $\beta c$ and bulk Lorentz factor $\Gamma = (1-\beta^2)^{-1/2}\gg 1$, this yields a temperature $T_{\rm eff}=\deltaD T_{\rm w} \sim E_{\rm p}/3k$ in the 
observer's frame.  Here $\deltaD = [\Gamma (1-\beta \cos\theta_{\rm obs})]^{-1}$ is the Doppler factor of the wind for an observer viewing it at 
an angle $\theta_{\rm obs}$ to its velocity.  Generally, the most intense emission arises when $\theta_{\rm obs}\lesssim 1/\Gamma$ is small and the wind is viewed head-on, corresponding to $\deltaD \sim \Gamma \gg 1$. The
energy density of the thermal radiation~\cite{Rybicki-1979} in the wind
frame is $U_{\rm th}= \pi^2/15\, (m_ec^2/\lambdac^3)\, \Theta_{\rm w}^4
\equiv aT_w^4$, for dimensionless temperature \teq{\Theta_{\rm w} =
kT_{\rm w}/m_ec^2}, where \teq{\lambdac = \hslash/m_ec} is the reduced
Compton wavelength of an electron, and $a=7.56 \times 10^{-15}$erg
cm$^{-3}$ K$^{-4}$ is the radiation constant\cite{Rybicki-1979}. The
corresponding pressure is \teq{P_{\rm th} = U_{\rm th}/3}, and its
enthalpy is \teq{W_{\rm th} = U_{\rm th} + P_{\rm th} = 4U_{\rm th}/3}.
Boosting to the star frame, the energy flux of the photons through a
surface locally perpendicular to this boost \cite{Blandford-1976-PhFl} is
${\cal F}_{\rm th}(\deltaD ) \approx \deltaD^2 W_{\rm th} c = 4\pi^2/45 \,
(m_ec^3/\lambdac^3)\, \deltaD^2\, \Theta_{\rm w}^4$, so that ${\cal F}_{\rm
th}(\deltaD ) \propto \deltaD^2$.  We choose bulk $\deltaD \sim \Gamma\sim 100$, as deduced from simple spherical dynamics arguments \cite{LAT2020}. 
Setting $\Theta_{\rm w} \to kT_{\rm eff}/(\deltaD m_ec^2)$ quickly yields
a thermal energy flux ${\cal F}_{\rm th} (\deltaD ) \approx 8 \times
10^{31} (10^2/\deltaD)^2$ erg cm$^{-2}$s$^{-1}$. This is clearly much
larger than the aforementioned flux inferred from observations, and much more so assuming that the radiation emanates from altitudes $R\gtrsim
10^9-10^{10}$cm where the plasma becomes transparent to electron
scattering. Adopting a smaller Doppler factor, $\deltaD \sim 10$, accommodating more closely the $\Gamma\sim 6$ bound obtained from the pair transparency considerations, increases the thermal flux and renders the disparity with the observed flux more extreme.  Thus, the thermalization in the wind is incomplete, consistent with the inherently non-Planckian form of the spectra at all times. Then, the radiation pressure exerted on the plasma is inferior to that invoked by applying traditional spherical GRB thermal fireball models \cite{Nakar-2005-ApJ}, possibly by factors of $\sim 10-10^4$.

The GBM emission is subjected to prolific Compton/Thomson scattering by electrons and positrons, which shape the non-thermal spectrum. The observer frame pair density $n_e$ of the wind couples to the radiation density $n_{\gamma}$ through a multi-layered interplay between expansion dynamics, geometry of flaring magnetic field lines that guide the outflow at altitudes below $10^8-10^9$cm, radiative transfer in a strongly-magnetized plasma, and magnetic opacity and pair equilibria in the wind frame. The complexities of these are beyond the scope of this paper. Yet, for a mean electron energy $\Gamma m_ec^2$ in the observer frame, one can connect $n_e$ to the emitted energy flux via a simple estimate $\epsilon n_e (\Gamma m_ec^2) \, c\sim {\cal F}(R) $ for some unknown radiative efficiency $\epsilon$ that is anticipated to not be vastly different from unity. Using Table~1 for interval (3) indicates that ${\cal F}(R)\sim 2 \times 10^{34} (10^6\hbox{cm}/R)^2$ erg cm$^{-2}$s$^{-1}$.  Assuming $\epsilon \sim 1$, one can discern that $n_e\sim 8 \times 10^{29}\,$cm$^{-3}$ at the stellar surface (where $\Gamma\sim 1$), and the optical depth $\taut = n_e\sigt R$ to Thomson scattering (of cross section  $\sigt = 6.65\times 10^{-25}$cm$^2$) is probably $\sim 5\times 10^{11}$ there.  \teq{\taut} doesn't drop below unity until altitudes of $R\gtrsim 10^{10}$cm, since $n_e\propto R^{-3}$, because magnetic field line colatitude \teq{\theta} and radius \teq{R} satisfy $\sin^2\theta \propto R$ in flaring dipolar field geometry. Magnetic reduction of the scattering opacity \cite{Paczynski-1992-AcA} can lower this transparency radius by a decade or so, and concomitantly decrease the radiation pressure and lower the ultimate bulk Lorentz factor somewhat.

A key GBM result is the exponentially decaying tail of the flux in Interval 4 (3-136.4~ms), after the initial spike (see Fig. \ref{fig:Fig2}). During this decay, which is of a timescale, $\tau \sim 45$ms, there is a distinctive ${\cal F}\propto E_{\rm p}^2$ correlation, depicted in  Fig. \ref{fig:Fig2}. This is the hallmark of a relativistic wind. The intrinsic couplings of observer frame flux ${\cal F}\propto \deltaD^2$ and peak energy $E_{\rm p}\propto \deltaD$ identified above, are true for a relativistic boost from the wind frame regardless of whether the photons are thermalized in the wind or not. Combined, they naturally generate the observed ${\cal F}\propto E_{\rm p}^2$ correlation at an approximately fixed emission radius. A distribution of Doppler factors $\deltaD$ is anticipated when sampling flare evolution as the observer angle $\theta_{\rm obs}$ changes significantly as the magnetar rotates, precipitating spread in the $E_{\rm p}$ and ${\cal F}$ distributions.  Flaring of the magnetic field lines when sampling different emission radii may modestly broaden these distributions.

The extremely short rise time of $T_{\rm rise}\sim 80\mu$s corresponds to a physical scale of $cT_{\rm rise}\sim 2.4\times 10^6$cm. The tail decay time is a factor of $\sim 550$ longer. These can potentially constrain emission region sizes that are vastly smaller than those inferred for GRBs from their variability times.  Yet this connection is impacted by the physical rotation of the magnetar. The tail decay timescale of $\tau =45\,$ms corresponds to a stellar rotation through an angle of $\Delta\theta \sim 2^\circ$ for a magnetar of typical period ($P=8$s). This type of duration is naturally expected for relativistic beaming of radiation from outflows with $\Gamma\sim 1/\Theta_{\rm coll} \sim 30$ for $\Theta_{\rm coll}\sim 2^{\circ}$. Even if the physical angular extent of the emission region exceeds the relativistic collimation angle $\Theta_{\rm coll}$, Doppler boosting dictates that the dominant signal will arise when the instantaneous observing angle to the wind velocity is of the order of $\theta_{\rm obs} \lesssim \Theta_{\rm coll}$. Thus, a picture emerges that what we might be seeing in the GBM, \emph{Swift}-BAT and KONUS-Wind signals is produced through stellar rotation of the Lorentz of Doppler-boosted emission in and out of our view. Accordingly, temporal scales $\Delta t \sim P/(2\pi \Gamma )$ may be signatures of the relativistic beaming structure associated with the GF wind convolved with the magnetar's rotation, what can be termed a relativistic lighthouse sweeping effect. The greater Doppler boosting near the $\theta_{\rm obs} \lesssim \Theta_{\rm coll}$ ``core'' of the Lorentz cone would yield harder spectra near the peak of intensity and softer in its ingress and egress, as is observed in Fig. \ref{fig:Fig1}.  

\secheader{Quasi-periodic Search and Analysis}
QPOs have been observed previously during three giant flaring episodes from 
SGRs 0526-66, 1806-20 and 1900+14 at frequencies ranging from 18--625~Hz \cite{Barat1983,Israel2005,StrohmayerWatts2005,WattsStrohmayer2006}. Of these, the QPO detections for SGRs 1806-20 and 1900+14 were restricted to the oscillating tails well after the conclusion of the initial spike.  These periodicities have been interpreted as signatures of torsional sub-surface seismic oscillations triggered by the cataclysmic rupturing of the crust seeding the GFs. 

We searched for a spin frequency over a range of $\nu$ = 0.02-50 Hz using both an unbinned and a logarithmically binned periodogram, in particular, we searched for signals with at least $p < 0.01$ (corrected for the number of frequencies and segments). We did not detect any signal that could be associated with stellar rotation.

We searched for QPOs in the NaI GBM data for GRB 200415A using the same methodology as that used to search for QPOs in the GBM data for SGR J0501+4516\cite{huppenkothen2013}, over a range of $\nu = 40-4000$~Hz. Due to telemetry packets dropping shortly after the peak, we searched three segments independently: the initial spike, $T_s = [-5, 5]$~ms, the fall time, $T_f = [7, 160]$~ms, and a long segment of $200\mathrm{s}$ after the initial spike to look for potential neutron star pulsation nominally in the range 0.05-1~Hz in the putative sub-background emission. We find no credible signal in the initial spike or the long segment. In the burst decay, we find a potential broad QPO candidate with moderate significance at $\nu \sim 180\, \mathrm{Hz}$. As the signal is broad, we use three different strategies to establish its significance. First we determine the trial-corrected $p$-value (probability) for a logarithmically-rebinned periodogram under the assumption of pure white noise is $p = 8.3 \times 10^{-4}$. However, the candidate's frequency places it at the edge of the frequency regime where stochastic variability in the form of red noise is important. To test whether the QPO can be explained by a combination of the overall decay of the light curve and the associated window function, we fit the light curve of this segment with an exponential function, and draw $1000$ sets of parameters from a multi-variate Gaussian using the best-fit solution and inverse Hessian derived from the optimization. We use these parameters to generate simulated light curves from the exponential function by adding white noise and generating periodograms in the same way we did for the data. The p-value for a QPO at that frequency under the assumption of an exponentially decaying light curve is $p = 0.023$, indicating that while less decisive, a decaying exponential cannot alone explain the observed excess in power at 180 Hz. We then use the methodology from Huppenkothen et al.\cite{huppenkothen2013} and model the periodogram with a power law (low-frequency red noise) and a constant (white noise), then search for outliers in the residuals. Using this method, we find a posterior-predictive p-value of $p = 0.016$, indicating that pure red noise cannot easily explain the excess power in the observed data. Finally, as the QPO is fairly broad, we model the periodogram with a combination of a power law, a Lorentzian and a constant, then compute the likelihood ratio between this model and the red noise-only model. Using the latter, we generate red noise via Markov-Chain Monte Carlo simulations using \textit{emcee} \cite{emcee} and derive a posterior predictive p-value for the likelihood ratio, $p = 0.031$. The parameters are well-constrained and uni-modal. We find a posterior centroid frequency for the QPO of $\nu = 183.2 \pm 19.9$ Hz, with a width of $\gamma = 46.7 \pm 29.7$ Hz. We used the posterior distribution of QPO parameters to derive the fractional root mean squared (RMS) amplitude, and find $A_\mathrm{frac} = 0.39 \pm 0.017$. Overall, while the significance is not consistently above a $3\sigma$ threshold, the significance remains consistently small in all attempts of quantification. 

We similarly searched for the same signal in data from the GBM BGO detectors, as well as BAT data and find no counterpart in either, which may be related to the lower count rates in the BAT, as well as a potential energy-dependence of the candidate. Using the aforementioned methods, we performed independent searches (both in the GBM and BAT data) at and around the QPO frequencies reported by ASIM\cite{ASIM2020}, but did not find any QPO candidates. 

\secheader{Radio Search}
Four observations of the Sculptor Galaxy were taken with the Karl G. Jansky Very Large Array (VLA) \cite{JVLA} under the project VLA/20A-438, 4.3 to 51.2 days post burst with the full 1-2 GHz L-Band. 3C48 was used for the band-pass and flux calibration and J0025-2602 was used for the complex gain calibration. 

The flagging, calibration, and imaging were done using the Common Astronomy Software Application (CASA) \cite{CASA}. Automated flagging was done using the tfcrop and rflag routines. After flagging and calibration, the spectral windows that were free of radio frequency interference were selected for imaging and four rounds of self-calibration. The baselines shorter than 2.5 kilo-wavelengths were excluded from the imaging process due to the bright diffuse emission of the Sculptor galaxy. The resulting images have a FoV diameter of $\sim1/3$ of a degree. The dates of each 11 minute observation, along with its accompanying RMS noise are listed in Table \ref{tab:Table2}. We note that due to the bright emission from the center of the galaxy, there are some artifacts from the VLA arms affecting the RMS noise in parts of the image by a factor of $\sim1.5$. Overall, the noise in these images is high due to the bright point sources in the image and the removal of short baselines due to the bright extended emission of the galaxy. A direct comparison between images, as well as image subtraction using the CASA immath task, shows no significant variable or transient emission in the field. 

\secheader{Data Availability}
Gamma-ray data from CGRO-BATSE, Swift-BAT and Fermi-GBM can all be found in public repositories. Catalogs of these data are provided as citations in both the main text and the Methods. The raw VLA data is publicly available. The calibrated VLA data and images are accessible to the reader per request. 

\secheader{Code Availability}
Standard software packages such as `rmfit' for GBM and `XSPEC' for other instruments, are available on-line. The codes used to determine the significance of the BGO photons, to construct the BAT TTE DRMs and to determine the rise-time, are available per reasonable request. The VLA data was analyzed with publicly available software (CASA). The entire procedure for detecting and quantifying the QPOs implemented in this paper, is publicly available in `Stingray'. The algorithm used to determine the pulse pile-up of the GBM data is available in Vandiver et al., (2013)~\cite{Vandiver}. The SwiMM code is not publicly available. However, response functions can be used to reproduce the spectral results in this study and are available upon reasonable request. 

\newpage

\section*{Acknowledgements}
The Fermi GBM Collaboration acknowledges the support of NASA in the United States under grant NNM11AA01A and of DRL in Germany. P.V. acknowledges support from NASA grant 80NSSC19K0595. A.T. and J.J.DL. thank Takanori Sakamoto for providing access to the Swift Mass Model. The authors thank the National Radio Astronomy Observatory, a facility of the National Science Foundation operated under a cooperative agreement by Associated Universities, Inc.. D.H. acknowledges support from the DIRAC Institute in the Department of Astronomy at the University of Washington. The DIRAC Institute is supported through generous gifts from the Charles and Lisa Simonyi Fund for Arts and Sciences and the Washington Research Foundation. J.J.DL acknowledges this material is based upon work supported by the National Science Foundation under Grants PHY-1708146 and PHY-1806854 and by the Institute for Gravitation and the Cosmos of the Pennsylvania State University.

\section*{Author contributions statement}

O.J.R. led the research effort. Authors O.J.R, P.V., M.G.B., M.S.B., C.K., E.G., A.T., J.L. and J.A.K. wrote the manuscript. Authors O.J.R, P.V., E.Bi., D.H., M.S.B., P.N.B, S.I.C, J.J.DL., J.A.K, D.K., A.T., G.Y., S.G. and R.H. all contributed to the data analysis leading to results in this paper. E.Bi. completed the first analysis of the event as she was Burst Advocate during the trigger time of GRB 200415A (GCN~27587). M.G.B. and P.V. led the interpretation of results. Specifically, O.J.R. P.V., E.Bi. and G.Y. contributed to the spectral analysis of the event. Additionally, P.V. worked on the time variability of GRB 200415A with P.N.B. and D.K.. D.K. performed the $T_{90}$ duration calculation. M.S.B. worked with P.N.B. on the data handling, specifically, addressing the band-width issue causing the data saturation in the GBM data. P.V and M.S.B analyzed the highest energy photon from GBM and also the pulse pile-up analysis. D.H performed the QPO analysis. J.A.K and A.T. provided the \emph{Swift}-BAT event data. J.J.DL. and A.T. ran the simulations and created the response files necessary to perform analysis of the Swift-BAT data, which in turn was performed by them and P.V.. E.G. contributed to the abstract with O.J.R., A.vdH., J.L. and S.I.C. all contributed to the radio search and write-up, with S.I.C. performing most of the VLA analysis. S.G. provided initial spectral analysis (e.g. Epeak and Flux correlations), and redshift estimates using his method for short GRBs. R.H. performed population analysis of GRBs with P.V.. C.W.H. provided feedback and helped steer the paper through the GBM internal review process. E.Bu helped put the result in the context of other short GRBs and performed chance likelihood calculations. He also helped organize the research effort of this source by other collaborating missions. All authors reviewed the manuscript.

\section*{Competing Interests}

The authors declare that they have no competing financial interests.
 
\section*{Correspondence}
Correspondence and requests for materials should be addressed to O.J.R.~(email: oroberts@usra.edu).

\newpage

\begin{figure*}[ht!]   
\begin{center}
\includegraphics[width=\textwidth,trim=0 0 0 0,clip]{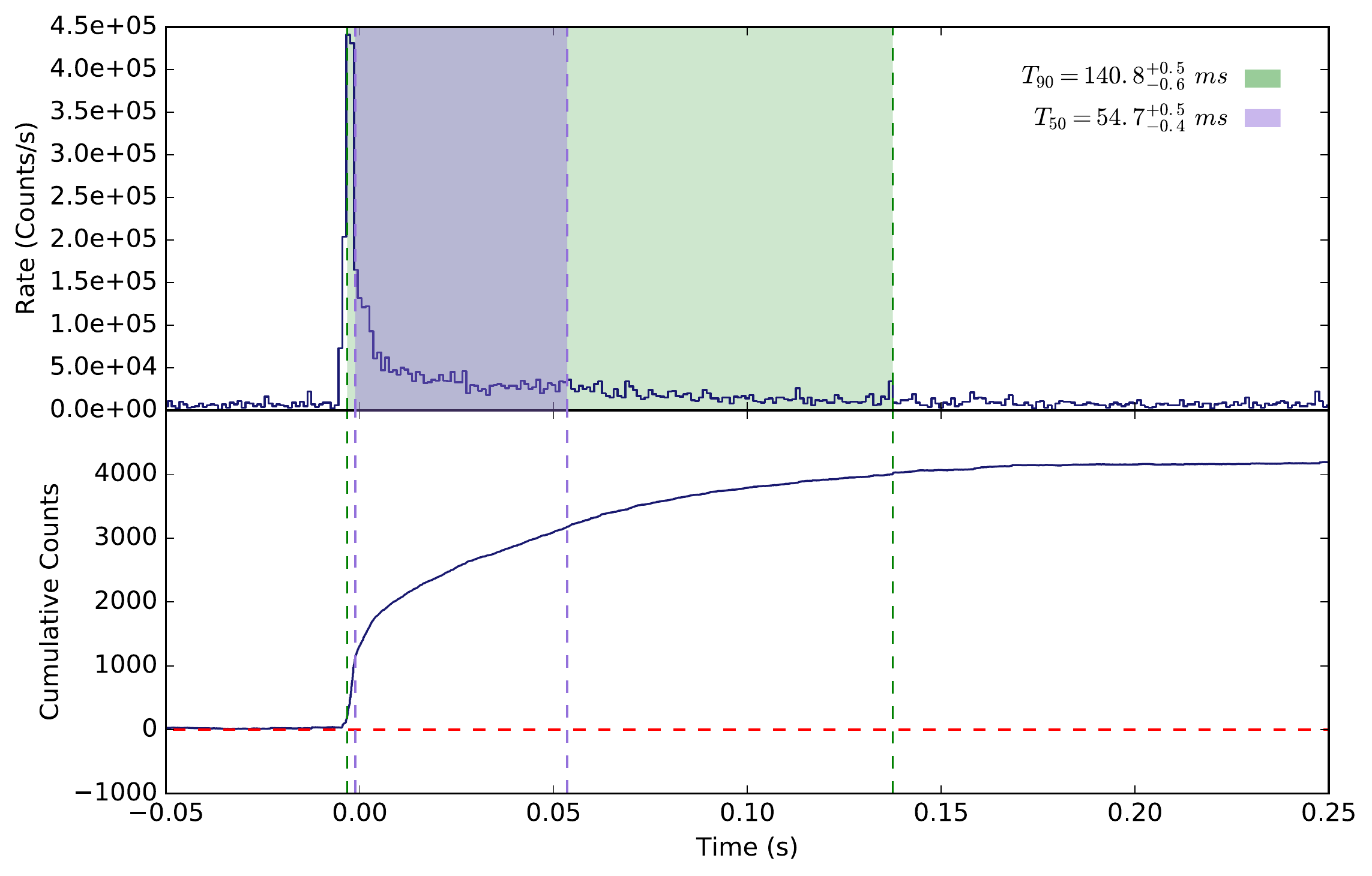}
\end{center}  
 ED Fig.~1. \textbf{The duration of GRB 200415A.} The $T_{90}$ (green) and $T_{50}$ (purple) durations, calculated using the \emph{Swift}-BAT data in counts-space. The errors are at the 1-$\sigma$ confidence level. \\
\label{fig:Fig3} 
\end{figure*}

\begin{figure*}[ht!]   
\begin{center}
\includegraphics[width=\textwidth,trim=0 0 0 0,clip]{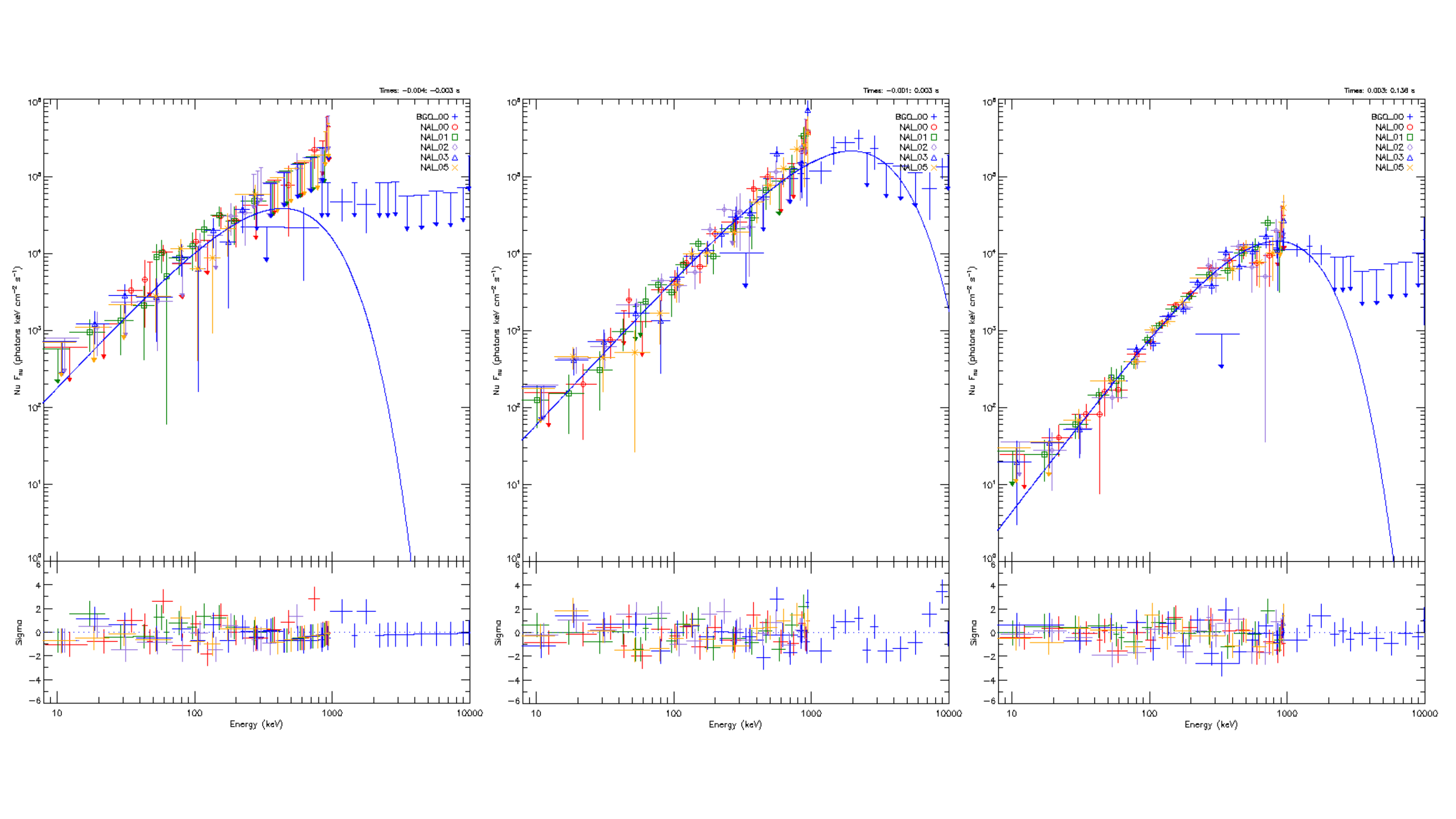}
 ED Fig.~2. \textbf{Spectra and fitted models in three time intervals for GRB 200415A.} $\nu F_{\nu}$ spectra and Comptonized fitting residuals for Interval 1 (Left), Interval 3 (Center) and Interval 4 (Right) of GRB 200415A. The three spectra are devoid of any instrumental effects attributed to band-width saturation, with the fit parameters listed in Table~\ref{tab:Table1}. These figures show the robustness of the fits to the data (1-$\sigma$ confidence) which are used in the main text and in Fig.~\ref{fig:Fig1}d, and are a direct result of the unrivalled temporal and spectral quality of the GBM data.\\
\label{fig:Fig4} 
\end{center}   
\end{figure*}


\begin{figure*}[ht!]   
\begin{center}
\includegraphics[width=\textwidth,trim=0 0 0 0,clip]{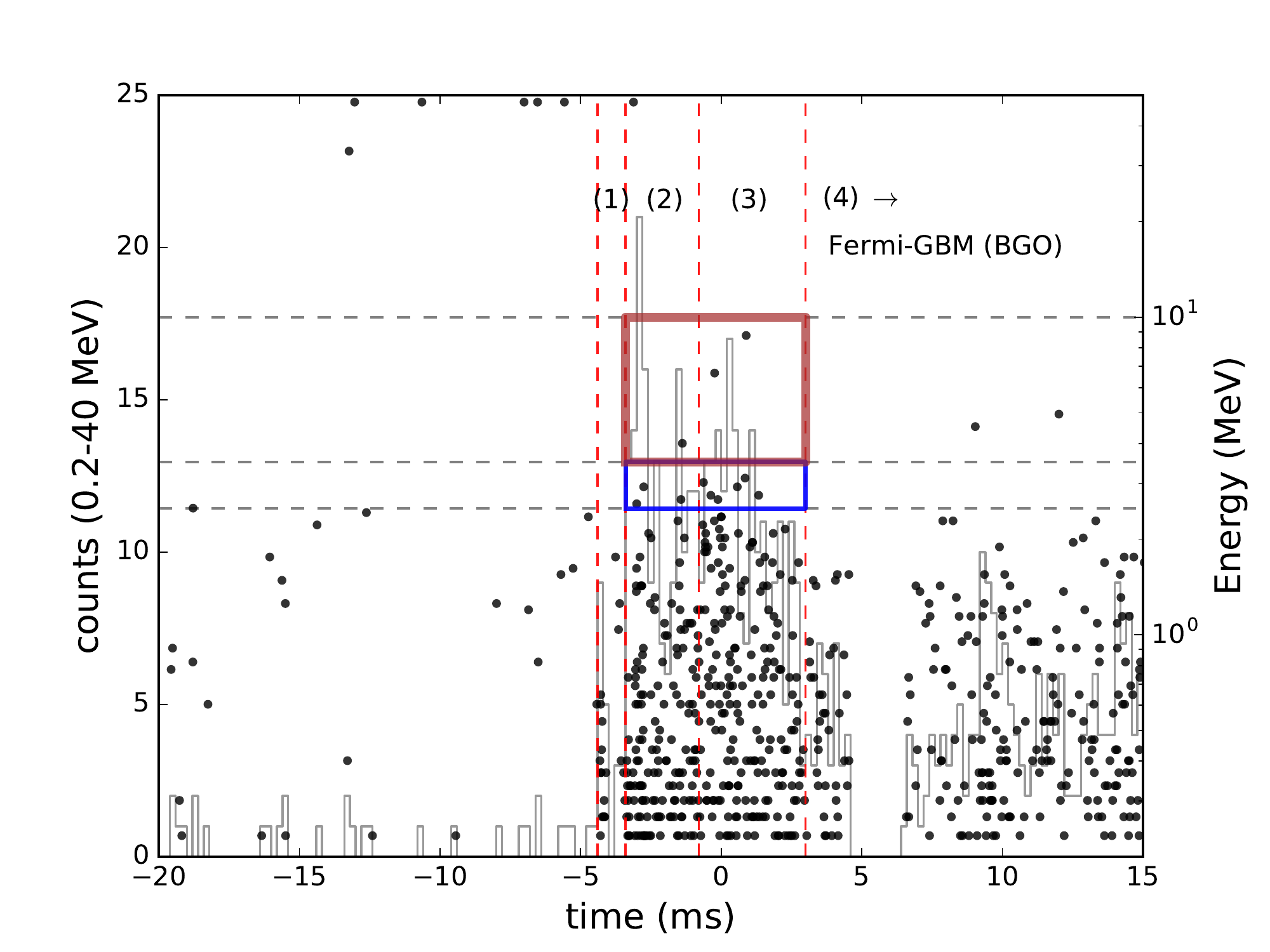}
ED Fig. 3. \textbf{Energetic photons from GRB 200415A.} Individual Time-Tagged Events of GBM BGO detector 0 (black dots). The blue rectangle indicates energies from 2.5 to 3.5 MeV in intervals (2) and (3), while the red rectangle shows energies from 3.5 to 10 MeV. We conclude that the highest  photon energy unambiguously associated to GRB 200415A is 3 MeV. 
\label{fig:Fig5} 
\end{center}   
\end{figure*}

\newpage

\begin{sidewaystable}
    \centering
	\caption{Spectral parameters, luminosity and emitted energy for the four time intervals identified in Fig.~\ref{fig:Fig1}a, relative to the GBM trigger time. The 1.896 factor corrects for the saturation in interval (2) by comparing the GBM flux in the 15-350 keV range with the Swift-BAT flux in the same interval. All errors are at the 1$\sigma$ confidence level. \label{tab:Table1}}
	\begin{tabular}{|c|c|c|c|c|c|c|c|}
		\hline
		Time  & $ E_{\rm peak}$ & photon index & Flux & Corr. & $L_{\gamma, iso}$& E$_{\gamma, iso}$ & C-Stat/dof\\
		(ms) & (keV) &  & ($10^{-5}$ erg cm$^{-2}$ s$^{-1})$ &  &  ($10^{47}$ erg s$^{-1}$)& ($10^{45}$ erg) &  \\ \hline
(1) $-4.4$ to $ -3.4$   &$428\pm71 $  &$-0.08 \pm 0.23$ & $9.9\pm1.2$       &1.0   &$ 1.51 \pm  0.18$ & $0.15\pm0.02$ & 431.0/633 \\
(2) $-3.4$ to $ -0.8$   &$997\pm77 $  &$-0.21\pm0.08 $ & $33.7\pm1.5$       &1.896 &$15.3  \pm  1.3 $ &$ 3.97 \pm  0.33$ &634.5/659 \\
(3)	$-0.8$ to $3.0$     &$1856\pm155$ &$-0.11 \pm 0.08$ & $17.5\pm 0.81$    &1.0   &$ 8.29 \pm  0.38$ &$ 3.15 \pm  0.15$ & 705.5/685 \\
(4)	$3.0$  to $136.4$   &$846\pm39$   &$ 0.34 \pm 0.08$ & $2.69\pm 0.06$    &1.036 &$ 0.58 \pm  0.032$&$ 7.79 \pm  0.43$ & 736.9/698 \\
	T$_{90}$ Duration (140.8) &  &     &  & & $1.07 \pm 0.17$ & $\mathbf{ 15.1 \pm 2.46}$ & \\
		\hline
	\end{tabular}
\end{sidewaystable}

\newpage
\begin{table}
    \centering
	\begin{tabular}{|l|c|c|}
		\hline
		Date & $\Delta T_{\rm{burst}}$ & 1$\sigma$ RMS noise \\
		(UTC) & (days) & (mJy/beam) \\ \hline
		April 19, 16:12:36 & 4.31 & 0.28 \\
		April 25, 16:26:53 & 10.3 & 0.43\\
		May 7, 15:04:53 & 42.3 & 0.40\\
		June 5, 12:43:29 & 51.2 & 0.29 \\
		\hline
	\end{tabular}
	\caption{1--2~GHz VLA radio observations}\label{tab:Table2} 
\end{table}
\end{document}